\documentclass[%
 aip,
 jmp,%
 amsmath,amssymb,
 reprint,%
 nofootinbib,%
 floatfix, 
]{revtex4-1}
\usepackage{graphicx}
\usepackage{caption}
\usepackage{subcaption}
\usepackage{dcolumn}
\usepackage{bm}
\usepackage[dvipsnames]{xcolor}
\usepackage{upgreek}
\usepackage{amssymb}
\usepackage{verbatim}
\usepackage{appendix}
\usepackage{listings}
\usepackage{units}
\usepackage{fancyhdr}
\usepackage{csquotes}
\usepackage{hyperref}
\usepackage{gensymb}
\usepackage[normalem]{ulem}

\usepackage{etoolbox}
\makeatletter
\patchcmd{\hyper@makecurrent}{%
    \ifx\Hy@param\Hy@chapterstring
        \let\Hy@param\Hy@chapapp
    \fi
}{%
    \iftoggle{inappendix}{
        \@checkappendixparam{chapter}%
        \@checkappendixparam{section}%
        \@checkappendixparam{subsection}%
        \@checkappendixparam{subsubsection}%
        \@checkappendixparam{paragraph}%
        \@checkappendixparam{subparagraph}%
    }{}%
}{}{\errmessage{failed to patch}}

\newcommand*{\@checkappendixparam}[1]{%
    \def\@checkappendixparamtmp{#1}%
    \ifx\Hy@param\@checkappendixparamtmp
        \let\Hy@param\Hy@appendixstring
    \fi
}
\makeatletter

\newtoggle{inappendix}
\togglefalse{inappendix}

\apptocmd{\appendix}{\toggletrue{inappendix}}{}{\errmessage{failed to patch}}
\apptocmd{\subappendices}{\toggletrue{inappendix}}{}{\errmessage{failed to patch}}

\newcommand*\obar[2][0.75]{
    \sbox{\myboxA}{$\m@th#2$}%
    \setbox\myboxB\null
    \ht\myboxB=\ht\myboxA%
    \dp\myboxB=\dp\myboxA%
    \wd\myboxB=#1\wd\myboxA
    \sbox\myboxB{$\m@th\overline{\copy\myboxB}$}
    \setlength\mylenA{\the\wd\myboxA}
    \addtolength\mylenA{-\the\wd\myboxB}%
    \ifdim\wd\myboxB<\wd\myboxA%
       \rlap{\hskip 0.5\mylenA\usebox\myboxB}{\usebox\myboxA}%
    \else
        \hskip -0.5\mylenA\rlap{\usebox\myboxA}{\hskip 0.5\mylenA\usebox\myboxB}%
    \fi}
\makeatother

\definecolor{Gold}{rgb}{1,0.84,0}

\definecolor{C1}{RGB}{51,34,136}
\definecolor{C2}{RGB}{136,204,238}
\definecolor{C3}{RGB}{68,170,153}
\definecolor{C4}{RGB}{17,119,51}
\definecolor{C5}{RGB}{153,153,51}
\definecolor{C6}{RGB}{221,204,119}
\definecolor{C7}{RGB}{102,17,0}
\definecolor{C8}{RGB}{204,102,119}
\definecolor{C9}{RGB}{136,34,85}
\definecolor{C10}{RGB}{170,68,153}

\definecolor{OC1}{RGB}{166,206,227}
\definecolor{OC2}{RGB}{31,120,180}
\definecolor{OC3}{RGB}{178,233,138}
\definecolor{OC4}{RGB}{51,160,44}
\definecolor{OC5}{RGB}{251,154,153}
\definecolor{OC6}{RGB}{227,26,28}
\definecolor{OC7}{RGB}{253,199,111}
\definecolor{OC8}{RGB}{255,127,0}
\definecolor{OC9}{RGB}{202,178,214}
\definecolor{OC10}{RGB}{106,61,154}

\definecolor{OOC1}{RGB}{195,170,60}
\definecolor{OOC2}{RGB}{93,54,134}
\definecolor{OOC3}{RGB}{101,188,103}
\definecolor{OOC4}{RGB}{194,106,187}
\definecolor{OOC5}{RGB}{112,143,57}
\definecolor{OOC6}{RGB}{108,126,215}
\definecolor{OOC7}{RGB}{182,120,55}
\definecolor{OOC8}{RGB}{70,193,154}
\definecolor{OOC9}{RGB}{185,74,115}
\definecolor{OOC10}{RGB}{186,76,65}

\definecolor{OOOC1}{RGB}{94,66,0}
\definecolor{OOOC2}{RGB}{71,103,222}
\definecolor{OOOC3}{RGB}{214,214,70}
\definecolor{OOOC4}{RGB}{85,0,71}
\definecolor{OOOC5}{RGB}{40,213,123}
\definecolor{OOOC6}{RGB}{220,46,88}
\definecolor{OOOC7}{RGB}{5,121,84}
\definecolor{OOOC8}{RGB}{255,169,246}
\definecolor{OOOC9}{RGB}{0,89,41}
\definecolor{OOOC10}{RGB}{250,143,56}

\newcommand{\remove}[1]{{}}

\newcommand{\appref}[1]{\hyperref[#1]{Appendix~\ref{#1}}}

\usepackage{morefloats}

\definecolor{hreflinkcolor}{rgb}{0.13,0.17,0.83}
\hypersetup{colorlinks=true,urlcolor=hreflinkcolor,
		linkcolor=hreflinkcolor,citecolor=hreflinkcolor}

\begin{document}

\preprint{AIP/123-QED}

\title{Proton acceleration by a pair of successive ultraintense femtosecond laser pulses}
\author{J.~Ferri}
\email{julienf@chalmers.se} 
\affiliation{Department of Physics, Chalmers University of Technology,
  SE-41296 G\"{o}teborg, Sweden}
\author{L.~Senje}
\affiliation{Department of Physics, Lund University, SE-22100, Lund, Sweden}
\author{M.~Dalui}
\affiliation{Department of Physics, Lund University, SE-22100, Lund, Sweden}
\author{K.~Svensson}
\affiliation{Department of Physics, Lund University, SE-22100, Lund, Sweden}
\author{B.~Aurand}
\affiliation{Department of Physics, Lund University, SE-22100, Lund, Sweden}
\author{M.~Hansson}
\affiliation{Department of Physics, Lund University, SE-22100, Lund, Sweden}
\author{A.~Persson}
\affiliation{Department of Physics, Lund University, SE-22100, Lund, Sweden}
 \author{O.~Lundh} 
\affiliation{Department of Physics, Lund University, SE-22100, Lund, Sweden}
\author{C.-G.~Wahlstr\"om} 
\affiliation{Department of Physics, Lund University, SE-22100, Lund, Sweden}
  \author{L.~Gremillet}
  \affiliation{CEA, DAM, DIF, F-91297 Arpajon, France}
\author{E.~Siminos} 
\affiliation{Department of Physics, University of Gothenburg,
  SE-41296 G\"{o}teborg, Sweden}
\author{T.~C.~DuBois} 
\affiliation{Department of Physics, Chalmers University of Technology,
  SE-41296 G\"{o}teborg, Sweden}
\author{L.~Yi} 
\affiliation{Department of Physics, Chalmers University of Technology,
  SE-41296 G\"{o}teborg, Sweden}
\author{J.~L.~Martins} 
\affiliation{Department of Physics, Chalmers University of Technology,
  SE-41296 G\"{o}teborg, Sweden}
\author{T.~F\"ul\"op} 
\affiliation{Department of Physics, Chalmers University of Technology,
  SE-41296 G\"{o}teborg, Sweden}

\date{\today}

\begin{abstract}
We investigate the target normal sheath acceleration of protons in thin aluminum targets irradiated at relativistic intensity by two time-separated ultrashort
($35\,\rm fs$) laser pulses. For identical laser pulses and target thicknesses of $3$ and $6\,\upmu\rm m$, we observe experimentally that the second pulse boosts the maximum energy and charge of the proton beam produced by the first pulse for time delays below $\sim 0.6-1\,\rm ps$.  By using two-dimensional particle-in-cell simulations we examine the variation of the proton energy spectra with respect to the time-delay between the two pulses. We demonstrate that the expansion of the target front surface caused by the first pulse significantly enhances the hot-electron generation by the second pulse arriving after a few hundreds of fs time delay. This enhancement, however, does not suffice to further accelerate the fastest protons driven by the first pulse once three-dimensional quenching effects have set in. This implies a limit to the maximum time delay that leads to proton energy enhancement, which we theoretically determine.
\end{abstract}


\maketitle

\section{Introduction}
\label{sec:intro}

The unique properties of laser-accelerated proton sources (e.g.~low emittance, short duration, high current) \cite{Cowan2004,Borghesi2004, Brambrink2006, Dromey2016} are at the basis of many well-established or potential applications \cite{Daido2012}, covering high-resolution probing of electromagnetic fields in plasmas \cite{Borghesi2004}, isochoric heating of dense materials \cite{Patel2003}, nuclear reactions \cite{McKenna2003}, isotope production \cite{Fritzler2003}, spallation physics \cite{McKenna2005} and tumor therapy \cite{Bulanov2002}. Several of the proposed applications, however, are very demanding with respect to particle energy and average flux. Optimization of the proton beam parameters can only be done through increased understanding of the fundamental acceleration processes.

The most studied laser-induced proton acceleration mechanism is target normal sheath acceleration (TNSA)\cite{Hatchett2000, Wilks2001, Mora2003, Passoni2010}, mainly due to its relatively low experimental requirements compared with other schemes (e.g.~radiation pressure acceleration or collisionless shock acceleration) \cite{Daido2012, Macchi2013}. In this mechanism, an ultraintense laser pulse interacts with a thin solid foil and heats the electrons to relativistic energies in the overdense plasma formed at the irradiated target surface. The expansion of the energized electrons into the vacuum induces a sheath electric field in the $\mathrm{TV m}^{-1}$ range, essentially normal to the target surfaces \cite{Snavely2000}. This sheath field can accelerate the protons from hydrogen-containing contaminants on the foil surface up to MeV energies over just a few micrometers. Depending on the laser-target parameters, efficient proton acceleration can occur at both the front and rear target sides \cite{Ceccotti2007}. Yet for a few-$\upmu\mathrm{m}$-thick foil and/or a not-so-high intensity contrast ($<10^{10}$) of the peak of the laser pulse compared to the pedestal, the fastest protons are mainly produced from the target rear surface \cite{Fuchs2005}. The maximum proton energy from TNSA has been observed to scale with the laser intensity as $E_{max} \propto I_0^\alpha$ with $\alpha \sim \tfrac{1}{2}-1$ depending on the laser pulse duration \cite{Zeil2010}, with a current experimental record of $85\,\mathrm{MeV}$ achieved using $\sim 200\,\mathrm{J}$ laser pulses focused to $\sim10^{20}\,\mathrm{Wcm}^{-2}$ on-target intensities \cite{Wagner2016}.

In order to gain additional control over the proton beam properties, several studies have recently addressed the case of TNSA induced by multiple laser pulses. For instance, the use of two successive pulses has been proposed as a means of modifying the shape of the proton energy distribution with the appearance of spectral peaks \cite{Robinson2007}. Experiments have also been performed using picosecond laser pulses, showing possible enhancement of the proton yield and maximum energy, provided the pulse parameters are accurately controlled \cite{Markey2010, Scott2012}.

In this paper, we further investigate the properties of the proton beams generated by two successive, ultraintense femtosecond laser pulses with varying temporal separation, incident on micrometer-thick aluminum targets. We perform experiments with precise control of the temporal separation between the two pulses for two different target thicknesses. To explain the experimental observations, we 
perform two-dimensional (2D) particle-in-cell (PIC) simulations using the \textsc{epoch} code \cite{Arber2015}, and find qualitative and quantitative agreement regarding the dependence of the maximum proton energies and numbers on the time delay between the pulses. The experimental results indicate that the acceleration process can be affected by the second pulse for time delays as long as $\sim 0.6\,\rm ps$ in $3\,\upmu\rm m$-thick targets and $\sim 1\,\rm ps$ in $6\,\upmu\rm m$-thick targets.

We show that the maximum time delay for effective two-pulse coupling roughly coincides with the occurrence of multidimensional effects that weaken the TNSA induced by the first pulse: when the fastest ions have moved a distance of the order of the transverse extent of the electric sheath field, a further input of hot electrons no longer significantly boosts their energy. Interestingly, the much increased hot-electron temperature and number produced by the second pulse does not therefore sufficiently compensate for the dropping efficiency of TNSA. The possibility of a two-stage proton acceleration process was considered previously in Ref.~\onlinecite{Markey2010}, using picosecond pulses and 1D PIC simulations to interpret the experimental results. In this work, the use of femtosecond laser pulses allows us to investigate shorter time delays, and hence to accurately delimit the coupling between the two pulses. To this aim, we shall concentrate on the dependence of the proton energy and number on the relative time delay between the two pulses.

The structure of the paper is as follows. In Section~\ref{sec:exp}, we first report on the experimental results obtained with two identical laser pulses. This is followed by a careful numerical analysis in Section~\ref{sec:num}, in which experimental results are explained in light of 2D PIC simulations, and by the derivation of a simple model for TNSA acceleration in thin foils in Sec.~\ref{sec:model}.
Finally, Section~\ref{sec:concl} summarizes our conclusions.

\section{Experimental results}
\label{sec:exp}

\begin{figure}[b]
\includegraphics[width=0.5\textwidth]{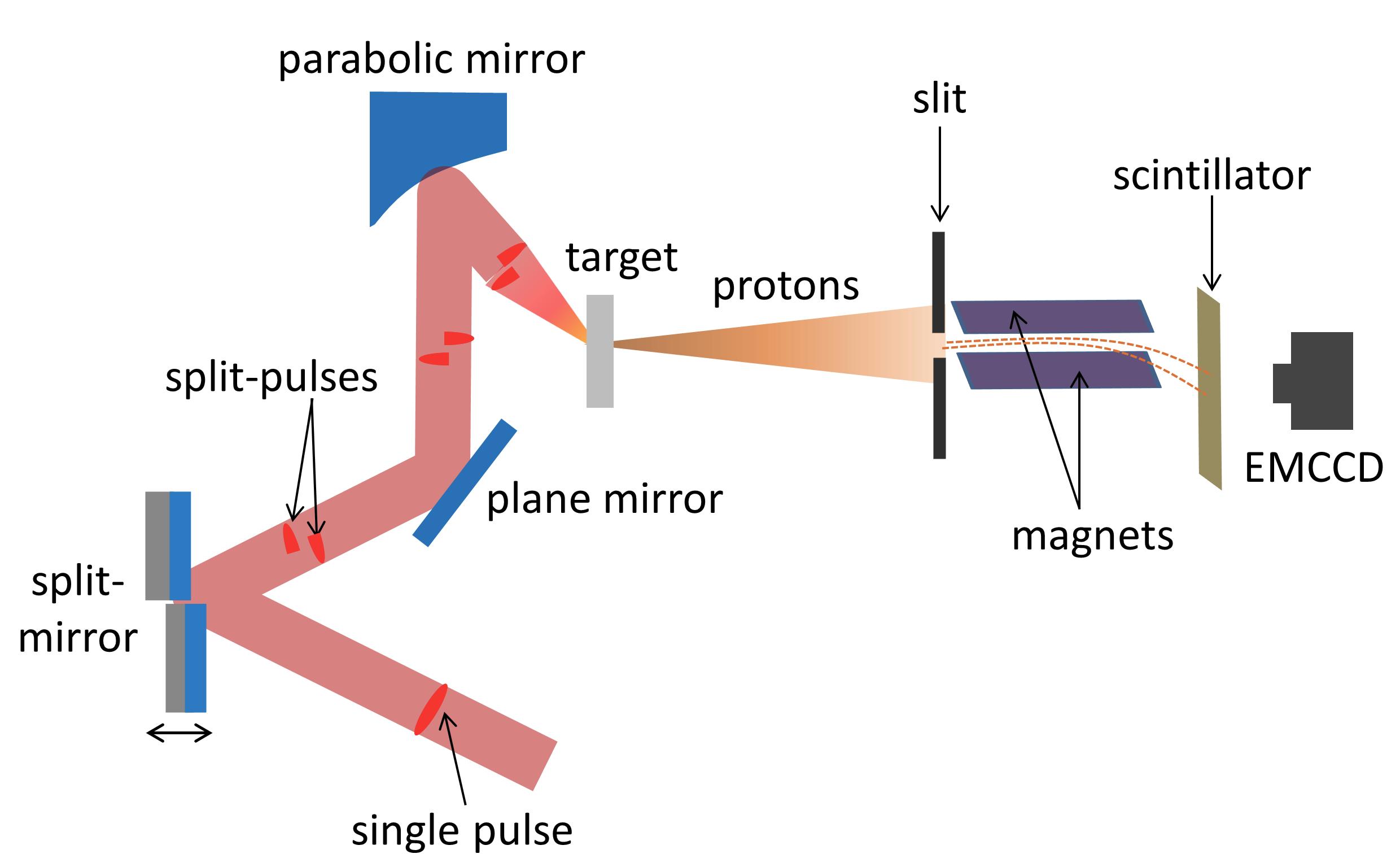}
\caption{\label{expt setup} Schematic of the experimental setup. The laser pulse is reflected on a split mirror where half of it is delayed with respect to the other half. Both the split pulses are focused by the same $f/3$ off-axis parabolic mirror onto a thin aluminum target foil, where they spatially overlap. The dimensions of the components and the separation of the pulses are heavily exaggerated for illustrative purposes.}
\end{figure}

\begin{figure}[b]
\includegraphics[width=0.5\textwidth]{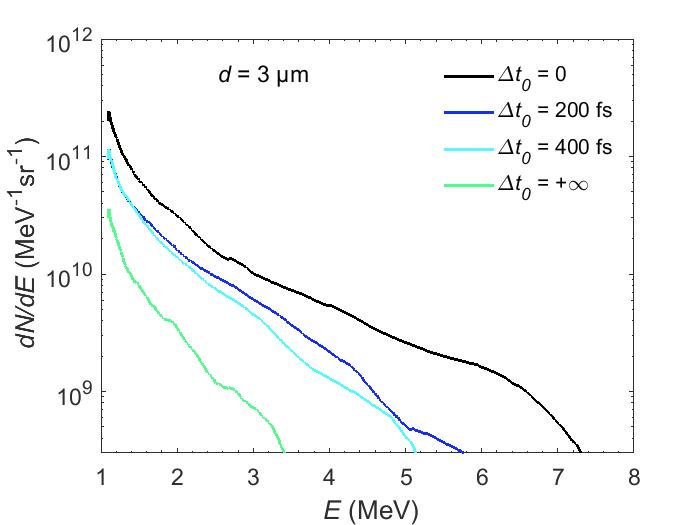}
\caption{\label{expt spectra} Experimentally measured proton energy spectra obtained from an Al target of thickness $d=3\,\upmu$m for different relative temporal delays between the two split pulses.}
\end{figure}

\begin{figure*}[t]
\includegraphics[width=1.0\textwidth]{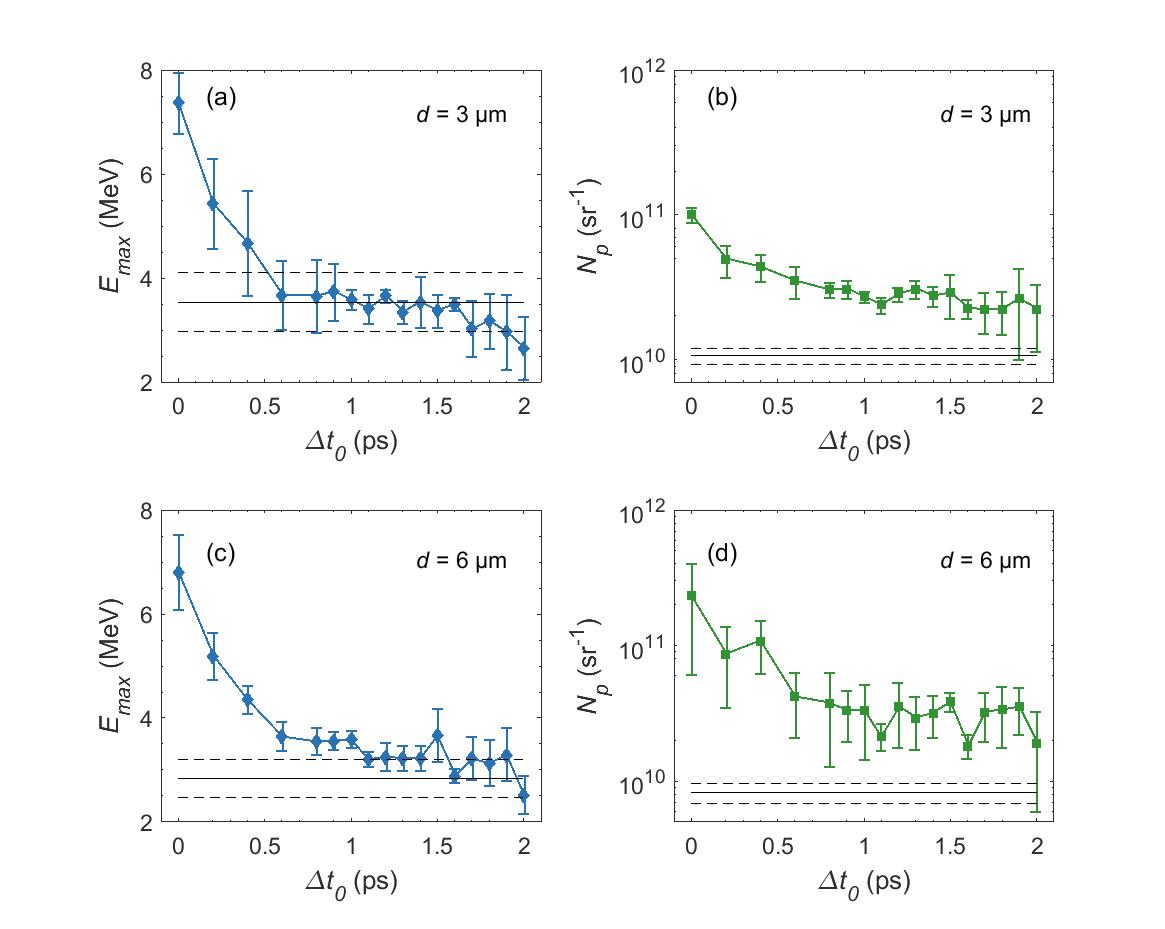}
\caption{\label{expt Emax and Np} Maximum proton energy (a,c) and proton number (b,d) for target thicknesses of $d=3\,\upmu\rm m$ (a,b) and $d=6\,\upmu\rm m$ (c,d). The mean value of 5 individual measurements for each delay is indicated by solid circles and rectangles; and the error bars show the corresponding standard deviation. The horizontal solid line corresponds to $\Delta t_0=\infty$ (the case of a single pulse of $0.4\,\rm J$ energy) and the dashed lines indicate the corresponding standard deviation.}
\end{figure*}

The experiments are performed using the multi-terawatt laser system at the Lund Laser Centre, which is a chirped pulse amplification based laser system and uses Ti:sapphire crystals as the active material for amplification. It delivers laser pulses of $35\,\rm fs$ duration and $0.8\,\upmu\rm m$ central wavelength, at a repetition rate of $10\,\rm Hz$. The intensity contrast of the pedestal, $100\,\rm ps$ prior to the peak of the pulse, is measured by a third-order auto-correlator to be $3\times 10^{9}$ during the experiment. Figure~\ref{expt setup} shows the schematic of the experimental setup. The laser pulse is split into two equal parts with a controllable delay between them using a split-mirror setup~\cite{Aurand2015}. The split-mirror consists of two plane mirrors placed very close to each other, with one mounted on a translational stage. This allows it to move in the perpendicular direction to its surface, enabling the time delay between the split pulses to be continuously varied up to a few ps. In addition, the whole assembly is mounted on another translational stage, which moves it transversely to the beam, allowing us to vary the energy ratio between the two pulses. In the present experiment, however, both split pulses deliver an energy of $0.4\,\rm J$, \textit{i.e.} half the total energy of the main pulse ($0.8\,\rm J$). 
Using a single $f/3$ off-axis parabolic mirror together with a deformable mirror that sends the main pulse onto the split-mirror, both pulses are focused and spatially overlapped in the focal plane. For perfect spatio-temporal overlap, the focal spot is measured to be $4\,\upmu\mathrm{m}\times 6\,\upmu\mathrm{m}$ (full width at half maximum), yielding a peak intensity of about $6\times 10^{19}\,\mathrm{Wcm}^{-2}$ in vacuum. Different targets consisting of aluminum foils of thicknesses of $3\,\upmu$m and $6\,\upmu\rm m$ are used. They are mounted on a target holder equipped with $x$-$y$-$z$ translational stages, and placed in the focal plane of the $p$-polarized laser beam at a $45^\circ$ angle of incidence. The accelerated protons are diagnosed by a magnetic spectrometer placed in the target-normal direction on the rear side of the target. The spectrometer consists of a $5\,\rm cm$-long permanent dipole magnet of $0.83\,\rm T$ field strength and a $1\,\rm mm$ thick plastic scintillator (BC-408) located at a distance of $10\,\rm cm$ behind the magnet. A $1\,\rm mm$-wide slit is used at the entry of the dipole magnet to sample the central part of the proton beam. In the spectrometer, the proton trajectories are bent according to their energy and then impinge on the scintillator. The resulting fluorescence is then imaged by a 16-bit EMCCD camera. The scintillator is covered with a $15\,\upmu\rm m$-thick aluminum foil, which stops protons up to $1\,\rm MeV$ as well as ions of lower charge-to-mass ratio.

Measurements are carried out by varying the relative delay ($\Delta t_0$) between the split pulses \cite{Senje2017}. Figure~\ref{expt spectra} shows the proton energy spectra, averaged over five consecutive laser shots, for different $\Delta t_0$ when a $3\,\upmu\rm m$-thick foil is used as the target. Here, $\Delta t_0=0$ corresponds to the case when both pulses fully overlap, so that they essentially constitute a single pulse of $0.8\,\rm J$ energy; and $\Delta t_0=+\infty$ means that only the first of the two split pulses reaches the target foil. The effect of changing the relative delay between the two split pulses on the maximum energy ($E_{max}$) and the total number ($N_p$) of the accelerated protons (of energies $>1\,\rm MeV$) are shown in Figs.~\ref{expt Emax and Np} for different target thicknesses $d$. As $\Delta t_0$ is increased, the maximum proton energy drops down to a constant value equal to the value achieved with a single pulse of $0.4\,\rm J$ energy: at $d=3\,\upmu\rm m$ (resp. $6\,\upmu\rm m$), this saturated value is measured to be $E_{max} \simeq 3.5 \pm 0.6\,\rm MeV$ (resp. $\simeq 2.8 \pm 0.4\,\rm MeV$) when $\Delta t_0 \gtrsim 0.6\,\rm ps$ (resp. $\gtrsim 1\,\rm ps$) [Figs.~\ref{expt Emax and Np}(a,c)]. The number of protons shows a similar trend as a function of $\Delta t_0$, except that, as observed up to $\Delta t_0 = 2\,\rm ps$, it seems to saturate above the value obtained with a single split pulse [Figs.~\ref{expt Emax and Np}(b,d)]. This indicates that, at large $\Delta t_0$, the second pulse may still contribute to increasing the number of $>1\,\rm MeV$ protons detected by the spectrometer, while it can no longer enhance their maximum energy.

According to Ref.~\onlinecite{Fuchs2007}, the time it takes for the most energetic protons to be accelerated through TNSA depends on the intensity and duration of the laser pulse. An often used rule of thumb is $t_{acc}=1.3(\tau_0+60)\,\rm fs$, with $\tau_0$ the laser duration, as inferred for laser intensities higher than $3\times 10^{19}\,\mathrm{Wcm}^{-2}$. Depositing more energy onto the target, in the form of a second laser pulse, after $t_{acc}$ or longer, might thus increase the number of fast protons, but should not affect their maximum energy. However, the acceleration time is a parameter introduced in the theoretical model to compensate for the decrease in hot electron temperature through energy transfer to the protons. It is thus interesting to investigate whether this acceleration time actually has any bearing in experimental studies. Assuming that the acceleration process is only affected by the second pulse while the process is still ongoing, it could then be possible to get information on the duration of the acceleration process by measuring how long the second laser pulse influences the maximum energy of the proton beam. However, our experimental results show that this occurs much more slowly than in the estimate given above, which yields $t_{acc} \simeq 120\,\rm fs$ for our experimental conditions. The maximum proton energy with an infinite relative delay can be used as a reference to determine when the second pulse no longer influences the acceleration process. It is clear that the proton energies are higher than this reference for much longer duration than $120\,\rm fs$ (about $0.6\,\rm ps$ to $1\,\rm ps$ depending on the target thickness). Therefore, a different interpretation will be pursued in the present work.

\section{PIC simulation results}
\label{sec:num}

To further understand the experimental results, we performed two-dimensional (2D) simulations with the \textsc{epoch} PIC code \cite{Arber2015}. In the simulations, two $p$-polarized laser pulses of $38\,\rm fs$ (FWHM) duration successively irradiate a solid-density Al foil at $45^\circ$ incidence angle. We use Gaussian pulses (both temporarily and spatially), with wavelength $\lambda_0=0.8\,\upmu\rm m$ and $0.4\,\rm J$ energy, which are focused to the same $5\,\mu\rm m$ spot size, leading to a peak on-target intensity $I_\mathrm{max} = 5\times 10^{19}\,\mathrm{Wcm}^{-2}$. When the relative time delay $\Delta t_0$ is zero, we use a single $0.8\,\mathrm{J}$ pulse. The dimensions of the simulation box are $66 \times 100\,\upmu\mathrm{m}^2$, with spatial steps $\Delta x = \Delta y = 10\,\rm nm$ ($6600 \times 10000$ cells). The target is composed of fully ionized $\mathrm{Al}^{13+}$ ions with initial density $n_i= 50n_c$ and electrons with density $n_e = 13n_i$.  Here, $n_c= m_e\varepsilon_0 \omega_0^2/e^2=1.742\times 10^{21}\,\mathrm{cm}^{-3}$ is the critical density, where $\omega_0=2\pi c/\lambda_0$ is the laser angular frequency, $m_e$ and $e$ are, respectively, the rest mass and charge of the electron, and $\varepsilon_0$ is the vacuum permittivity. We employ the physical ion-to-electron mass ratio $m_i/m_e = 1836\times 27$. To model the hydrogen-rich surface contaminants, the front and rear target sides are coated with electron-proton layers of $20\,\rm nm$ thickness and $100n_c$ density.  All plasma species have an initial temperature of $200\,\rm eV$ and are modeled using 50 macroparticles per cell, except for the protons for which 1000 macroparticles are used per cell for better statistics.

\begin{figure}
\includegraphics[width=0.5\textwidth]{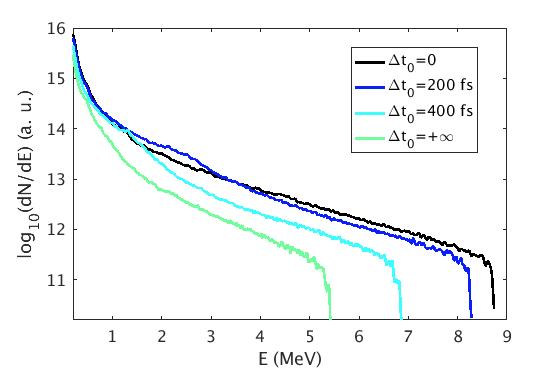}
\caption{\label{rear} Simulated energy spectra $dN_p/dE$ for the protons originated from the rear side of the target as obtained for
different time delays.  The spectra are plotted at time $t=1\,\mathrm{ps}$ (the peak of the first sub-pulse reaches the target at
$t=150\,\mathrm{fs}$).}
\end{figure}

\begin{figure}
\includegraphics[width=0.5\textwidth]{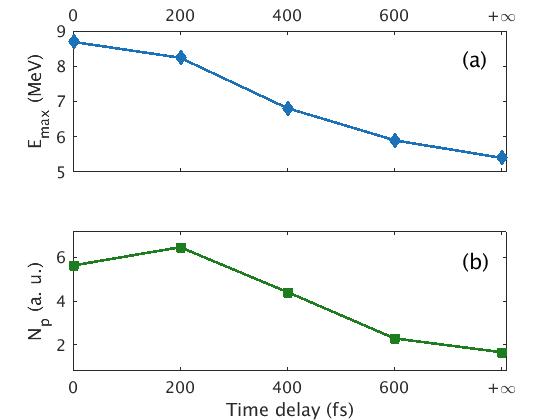}
\caption{\label{rear1} Dependence of (a) the maximum energy and (b) number ($N_p$) of rear-side protons with energies above $1\,\rm MeV$ on the time delay between the two pulses in the simulations of the $3\,\upmu\rm m$ target. All quantities are recorded at $t=1\,\rm ps$.}
\end{figure}

We simulated cases corresponding to time delays $\Delta t_0 = 0, 200, 400, 600\,\rm fs$ and $+\infty$ (a single laser pulse with $0.4\,\rm J$ of energy). Our simulations neglect preplasma formation, \textit{i.e.}~we assume an ultra-high temporal contrast. The front side of the Al target is situated at $x=0$, and the peak of the first sub-pulse arrives on the target at $t = 150\,\rm fs$. The total simulation time is $1\,\rm ps$. In the following, we will only address the acceleration of rear-side protons, even though front-side protons can also be driven to significant energies. Furthermore, we will mostly consider the $3\,\upmu\rm m$-thick target, although results for the $6\,\upmu\rm m$ target will be shown for comparison.

The simulated energy spectra of the protons initially located at the rear of the $3\,\mu\rm m$ target are shown in Fig.~\ref{rear}. These broadly dispersed spectra reproduce well the experimental results of Fig.~\ref{expt spectra}, both quantitatively and qualitatively, with the maximum energy, $E_{max}$, decreasing from $\simeq 8.6\,\rm MeV$ for $\Delta t_0=0$ to $\simeq 5.4\,\rm MeV$ for $\Delta t_0=+\infty$. The small overestimation of $E_{max}$ compared to the experiment is expected to be an effect of the 2D geometry in the simulations, which inaccurately describes the transverse expansion of the hot electrons and of the associated sheath fields. The cases $\Delta t_0=200\,\rm fs$ and $\Delta t_0 = 400\,\rm fs$ exhibit small breaks in the slope of the spectra, due to the modification of the acceleration process induced by the second pulse, as will be discussed later. We, however, do not observe the generation of spectral peaks, as was suggested in previous work with 1D simulations \cite{Robinson2007}. 

\begin{figure}[b]
\includegraphics[width=0.5\textwidth]{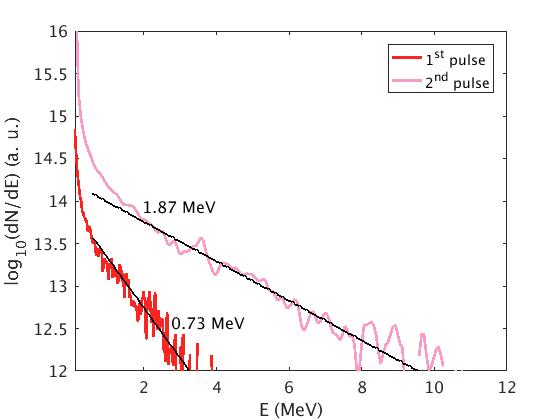}
\caption{\label{Electronspectra} Electron energy spectra $dN/dE$ after the arrival of the first half-pulse (red, $160\,\rm fs$), and after the arrival of the second half-pulse, for a $200\,\rm fs$ delay (pink, $360\,\rm fs$). Black lines correspond to Maxwellian fits for the spectra.}
\end{figure}

Figure~\ref{rear1} displays the dependence of the maximum energy and number ($N_p$) of protons with energies over $1\,\rm MeV$ on the time delay between the pulses. In agreement with the measurements, we find that the second pulse can affect the maximum proton energy for a time delay as long as $\Delta t_0 \simeq 600\,\rm fs$. For this time delay, we observe only a slight increase in the proton energy compared to the case with $\Delta t_0=+\infty$ (about 10~\%), whereas the number of protons is about twice the value for this case. The increase in $N_p$ for $\Delta t_0=200\,\rm fs$ points to a higher number of relatively low energy protons generated by the second pulse (as can be seen in Fig.~\ref{rear}), yet also depends on the chosen lower cutoff energy. 

In order to understand why the second pulse is important after such a time delay, we plot in Fig.~\ref{Electronspectra} the electron energy spectra right after the interaction of the peak of each laser pulse with the target (for the case $\Delta t_0=200\,\rm fs$), \textit{i.e.}, when the hottest electrons are produced. The temperature of the hot-electron distribution (fitted to a Maxwellian) is a factor of 2.5 higher for the second pulse, which demonstrates its greater effectiveness in producing hot electrons.

\begin{figure}
\includegraphics[width=0.5\textwidth]{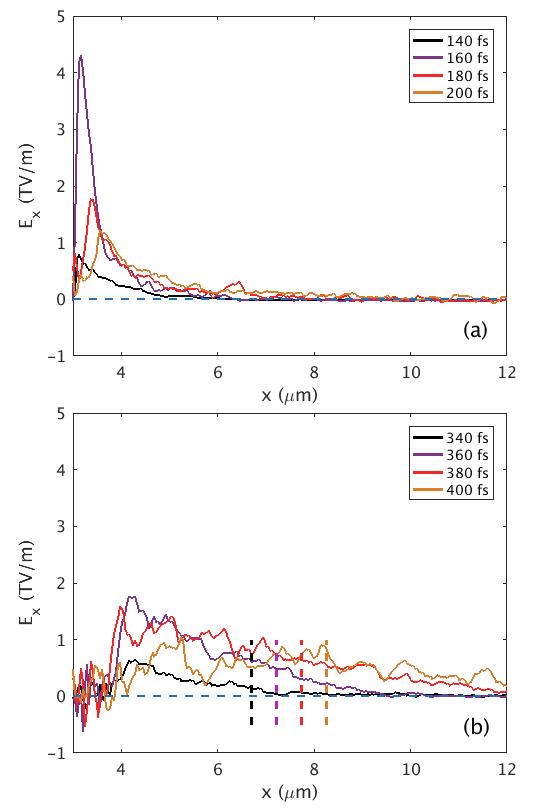}
\caption{\label{Ex} On-axis $E_x$ field (normal to the target rear surface where it is hit by the laser) at different times after the arrival (a) of the first sub-pulse and (b) of the second sub-pulse for a $200\,\rm fs$ time delay. Vertical dashed lines in (b) indicate the position of the proton fronts at the different times.}
\end{figure}

\begin{figure*}
\includegraphics[width=\textwidth]{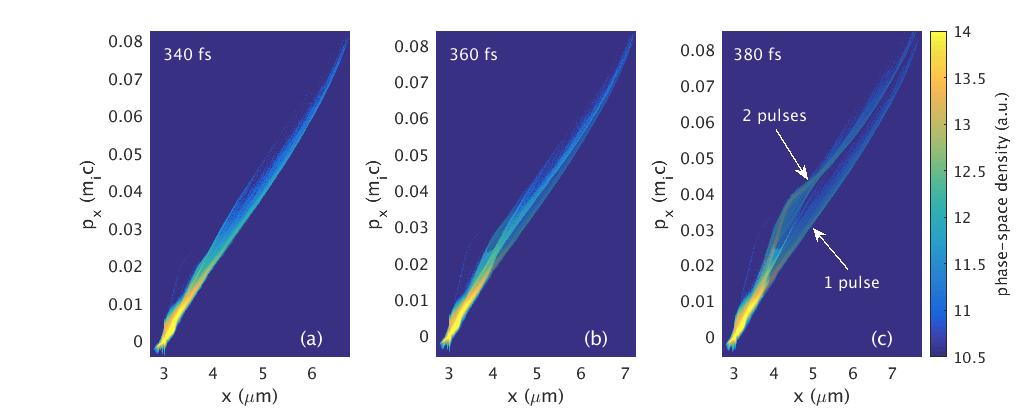}
\caption{\label{Qxpx} Longitudinal $x-p_x$ phase spaces of the rear-side protons at successive times. In each panel the phase spaces obtained with $\Delta t_0 = +\infty$ (1 pulse) and $\Delta t_0 = 200\,\rm fs$ (2 pulses) are superimposed. It is clear that the most energetic protons are those at the outermost front in both cases (compare the location of the proton beam with the accelerating field profile plotted in Fig.~\ref{Ex}b).}
\end{figure*}

We will elaborate later on the varying electron heating efficiency by the two subsequent pulses. First we show that the accelerating sheath electric field $E_x$ induced at the target rear will also be different for the two sub-pulses, as can be seen in Fig.~\ref{Ex}. Following the interaction with the first sub-pulse, $E_x$ reaches a maximum of $\sim 4\,\mathrm{TV m}^{-1}$, but quickly decreases both on short spatial and short time scales (a few tens of fs). The peak value of $E_x$ is consistent with the standard estimate\cite{Mora2003} $E_x \simeq \sqrt{n_h k_\mathrm{B} T_h/\varepsilon_0}$, where $T_h \simeq 0.7\,\rm MeV$ and $n_h \simeq n_c$ are respectively the hot-electron temperature and density at the target rear from the simulation. The sheath field strength increases again after the second pulse irradiation, yet reaches a smaller peak value ($\lesssim 2\,\mathrm{TV m}^{-1}$) than after the first pulse. The reason is that the target rear side has expanded over a few micrometers distance, much larger than the Debye length associated with the hot electrons, so the sheath field then scales as $E_x \sim T_h/eL_n$, with $L_n$ the density scale length \cite{Wilks2001}. Due to the plasma expansion and the higher flux of incoming hot electrons with higher energies, the sheath field extends further spatially and recedes more slowly temporally. This re-intensified field is expected to boost the energy of the protons being accelerated in the first-pulse-driven sheath field. However, as further explained below, the fastest protons (\textit{i.e.}, those located at the expanding plasma front) can benefit from this enhanced field only insofar as they are not too far away from the target rear side. In the present case of $\Delta t_0 = 200\,\rm fs$, the fast protons driven by the first pulse, which extend up to $x \simeq 7\,\upmu\rm m$ at the second laser peak (see Fig.~\ref{Qxpx}), indeed turn out to be re-accelerated. This leads to the bump in the proton $x-p_x$ phase space visible at $t=380\,\rm fs$. Note that due to transverse sheath expansion in 2D simulations, we never observe the proton spectral peaks that develop in 1D simulations. These observations remain true when using a time delay $\Delta t_0 = 400\,\rm fs$ or longer. 


\begin{figure}
\includegraphics[width=0.5\textwidth]{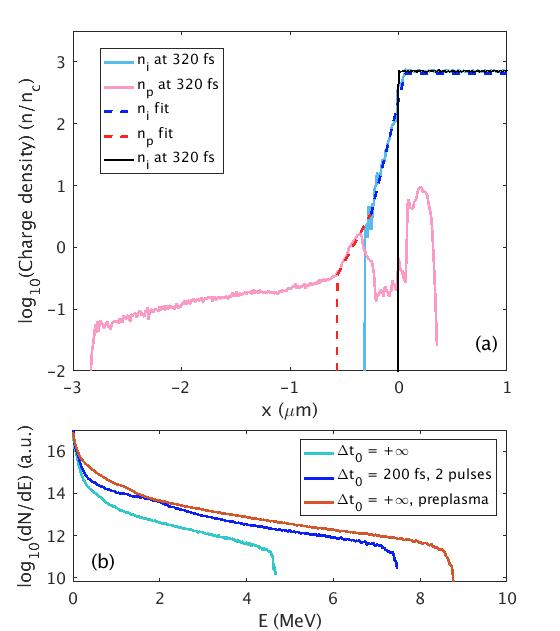}
\caption{\label{density} (a) Charge density profiles (normalized to $n_c$) for the various plasma species after $320\,\mathrm{fs}$ in the simulation with $\Delta t_0 = 200\,\mathrm{fs}$ (corresponding to the time when the second sub-pulse interacts with the target): ions (solid light blue) and protons on the front side (solid pink) and density profile at the beginning of the simulation with preplasma (dashed blue and red lines). The fit for the proton density does not take into account the low-density part ($<0.2n_c$) and the protons pushed into the target due to hole boring. The initial ion density without preplasma is shown in dark.
(b) Energy spectra for the protons on the rear side of the target after $700\,\rm fs$ of simulation. The results with the initial preplasma taken from the simulations (orange) are a proof-of-principle of the role played by a density gradient on the front side.}
\end{figure}

Our simulations indicate, in particular, that the second pulse benefits from better absorption as it produces a larger number of hotter electrons. This difference in behaviour is obviously caused by a change in the target state as seen by the two pulses, namely by the generation of a preplasma following the first pulse irradiation. When the second pulse interacts with the target, a $\sim 0.5~\upmu\rm m$-long plasma gradient has indeed formed on the front side, as can be seen in Fig.~\ref{density}. Such a micron-scale expansion of the target front is expected to increase the absorption of the second pulse \cite{Lefebvre1997, Mckenna2008, Nuter2008, Paradkar2011, Arefiev2015}. To support this prediction under the present conditions, we performed a simulation in which only one $0.4\,\rm J$ laser pulse impinges onto a $3\,\upmu\rm m$ target that comprises a front-side density profile similar to that seen by the second pulse after $200\,\rm fs$. As shown in Fig.~\ref{density}(a), we use exponential fits for the density profile for $n_i$ and $n_p$, and neglect the low-density part ($<0.2n_c$) of the preplasma, as well as those protons being driven into the target by the laser-induced hole boring. We also used the same initial temperature of $200\,\rm eV$ for the different species, meaning that the enhanced hot-electron generation by the second sub-pulse is assumed to result from the modified density profile (rather than from the re-acceleration of recirculating hot electrons). Note that we only modified the front of the target, while we kept a sharp rear boundary. 


As expected, the presence of a preplasma strongly enhances laser absorption and hot-electron generation, yielding a hot-electron spectrum similar to that observed in the simulation following the second sub-pulse (not shown). The accelerated protons therefore reach significantly higher energies than with a sharp target-vacuum interface [Fig.~\ref{density}(b)]. It is worth noting that the proton beam generated in this case is also slightly more energetic and of higher charge than the one generated using two sub-pulses with a $200\,\rm fs$ delay, although it ignores initial proton acceleration by the first sub-pulse. This stems from the fact that, in the two-pulse setup, the target rear side may significantly expand after the first pulse, which, as already shown (Fig.~\ref{Ex}), tends to weaken the sheath field driven by the second pulse. This feature is, however, not taken into account in this simulation (as we were only interested in the role of the front-side preplasma), but incidentally results in weaker proton acceleration \cite{Grismayer2006, Fuchs2007PRL}. Therefore, it proves more efficient to employ a single pulse with an increased preplasma size while keeping a sharp gradient on the rear of the target.

\begin{figure}
  \includegraphics[width=0.5\textwidth]{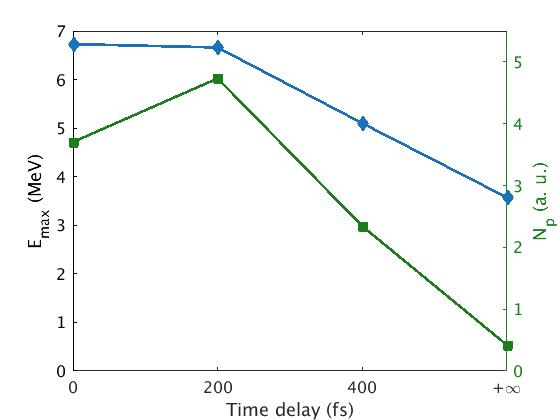}
  \caption{\label{6micron} Maximum proton energy (blue) and number of rear-side protons with energy above 1~MeV (green) after acceleration by a pair of identical laser pulses with
  varying temporal delay in the case of the $6\,\upmu\mathrm{m}$ target (values at $t=1.2\,\mathrm{ps}$).}
\end{figure}

Finally for the $6\,\upmu\rm m$-thick target, the PIC results also yield qualitative agreement with the experimental data. For this target, the simulation time is extended to 1.2~ps. Figure~\ref{6micron} shows that the 2D simulations reproduce the drop in the cutoff proton energies and in the number of protons that is observed in the experimental values. It shows a slight overestimation of the maximum proton energy $E_{max}$ compared with the experiment, but this dimensional effect is predicted by the model developed in Sect.~\ref{sec:model}.

\section{A simple model for thin-foil expansion}\label{sec:model}

The experimental results which show a steady decrease in maximal protons energies with increasing time delay can be explained by both a modified density profile at the target front between the arrival of the two sub-pulses and, more importantly, by an increasing distance of the highest energy protons from the rear of the target for longer time delays. An estimate of the maximum time delay, $\Delta t_{0,max}$, for which the second pulse can still boost the maximum proton energy can be obtained from the following simple model for ion expansion from a thin solid foil.

In the early stage of expansion, one can assume a 1D plasma dynamics and isothermal hot electrons. Under these conditions, the velocity of the fastest protons evolves as\cite{Mora2003}
\begin{equation}
  v_{f,is}(t) \simeq  2c_{s0} \ln \left( \sqrt{\omega_{p0}^2t^2/2e_N+1} + \omega_{p0}t/\sqrt{2e_N} \right) \,,
\end{equation}
where $c_{s0}=\sqrt{T_{h0}/m_p}$ is the initial (prior to the proton expansion) sound speed in the proton plasma,  $e_N=2.71828$ is Euler's constant, $\tau = \omega_{p0}t$ with $\omega_{p0} = \sqrt{n_{hr0} e^2/ m_p \varepsilon_0}$ the initial proton plasma frequency, and $n_{hr0}$ is the initial hot-electron density at the target rear side. The above expression also holds in the presence of a dense population of colder electrons of initial temperature $T_{c0}$, provided \cite{Mora2005} $T_{h0}/T_{c0} \gtrsim 10$, which is well fulfilled in the present conditions.

Due to the finite target thickness, the assumption of isothermal hot electrons should break down at a time $t_{ad} \simeq d/2c_s \simeq 180-360\,\rm fs$ in our parameter range. For $t>t_{ad}$, the electrons experience adiabatic cooling, and the ion acceleration rapidly slows down. The maximum proton velocity achievable in this adiabatic regime is approximately given by \cite{Mora2005}
\begin{equation}
    v_{f,ad} \simeq 2 c_{s0} \ln \left(0.32 d/\lambda_{D0}+4.2 \right) \,,
\end{equation}
with $\lambda_{D0} = (\varepsilon_0 k_B T_{h0}/n_{hr0})^{1/2}$ the initial Debye length. A simple formula for the maximum proton velocity, which smoothly transitions from the isothermal to the adiabatic expansion regimes, is given by
\begin{equation} \label{eq:vp_fit}
  v_{f,1D}(t) \simeq \left[v_{f,is}^{-2}(t) + v_{f,ad}^{-2}\right]^{-1/2} \,.
\end{equation}
From this expression readily follows the electric field strength seen by the fastest protons:
\begin{equation}\label{eq:Ex1D}
  E_{x,1D}(t) = \frac{m_p \dot{v}_{f,is}(t)}{2e\left[1+v_{f,is}^2(t)/v_{f,ad}^2\right]^{3/2}} \,,
\end{equation}
with $\dot{v}_{f,is}(t)=2c_{s0}\omega_{p0}/\sqrt{\omega_{p0}^2t^2+e_N}$.

The above formulas assume a 1D expansion geometry, which ceases to be valid when the fastest protons have moved a distance comparable with the transverse size of the sheath field $D_\perp \simeq w_L+2d \tan (\theta_h)$, with $w_L$ the laser spot size and $\theta_h$ the half-angle divergence of the hot electrons. An accurate analytic modelling of TNSA in a multidimensional geometry is a difficult, as-yet-unsolved problem. Here, drawing upon Ref.~\onlinecite{Brantov2015}, we limit ourselves to including a space-dependent factor in Eq.~\eqref{eq:Ex1D} that heuristically describes the expected fast decay of the sheath field once 2D or 3D effects set in. Specifically, we consider the following modified expression for the sheath field:
\begin{equation}\label{eq:Ex23D}
    E_x(t) = \frac{E_{x,1D}(t)}{\left[1+x_f^2(t)/D_\perp^2 \right]^{\frac{\delta -1}{2}}} \,,
\end{equation}
where $x_f(t)$ is the longitudinal position of the proton front, and $\delta \in (2,3)$ is the spatial dimensionality of the problem. The complete motion of the fastest protons is then obtained by numerically solving the coupled equations
\begin{align}
    \dot{x}_f(t) &= v_f(t)  \,, \\
    \dot{v}_f(t) &= \frac{e}{m_p}E_x(t) \,,
\end{align}
with the initial conditions $x_f(0)=v_f(0)=0$. This calculation requires the knowledge of the initial backside hot-electron density, which is taken to be $n_{hr0} = n_{h0}\left(w_L/D_\perp\right)^{\delta-1}$, with $n_{h0}$ the hot-electron density at the laser-irradiated target front.

\begin{figure}
\includegraphics[width=0.5\textwidth]{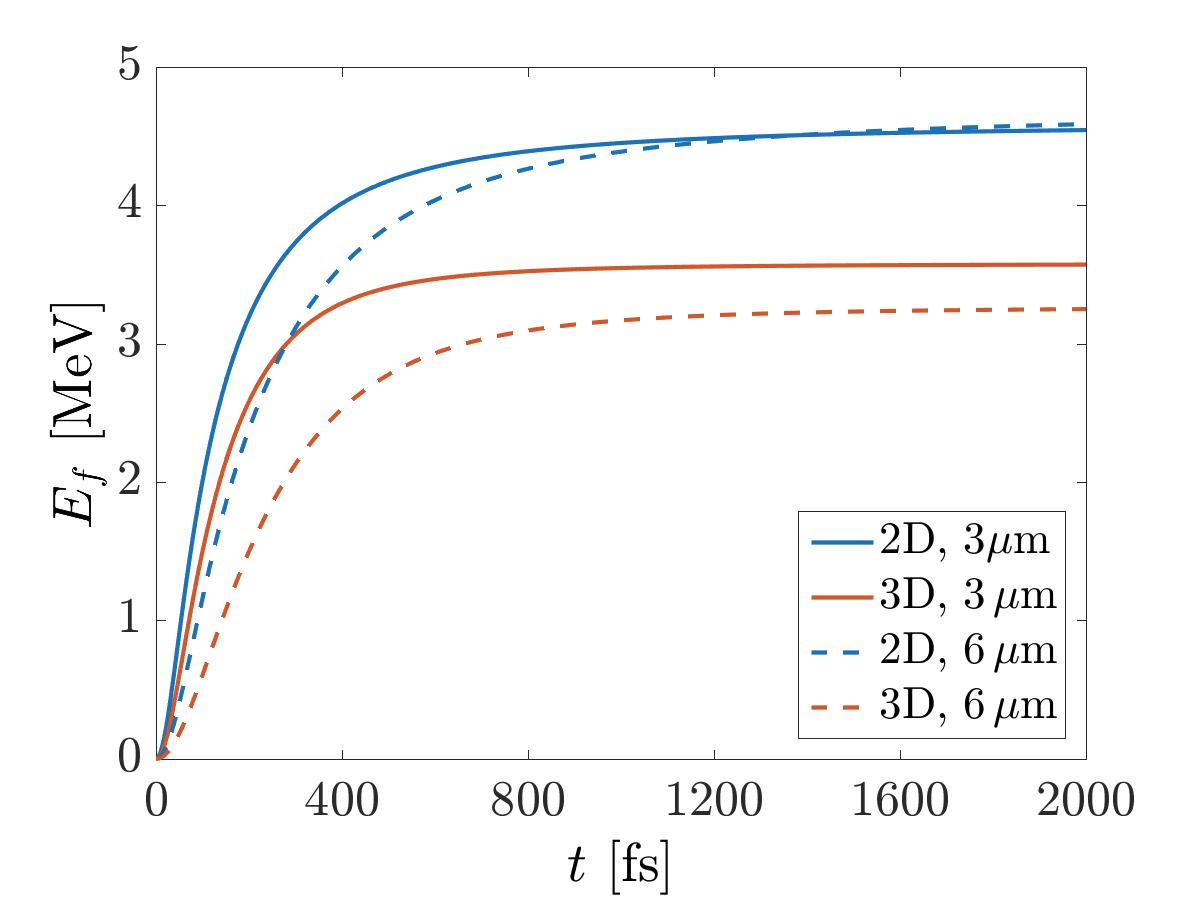}
\caption{Time history of the proton energy at the expanding plasma front for target thicknesses $d=3\,\upmu\rm m$ (solid) and $d=6\,\upmu\rm m$ (dashed), in 2D (blue) and 3D (red) geometries.
See text for details on the initial parameters.
\label{fig:model_ef}}
\end{figure}

Figure~\ref{fig:model_ef} plots the time evolution of the cutoff proton energies, $E_f(t) = m_p v_f^2(t)/2$, as predicted by the model for different geometries (2D, 3D) and target thicknesses ($3\,\upmu\mathrm{m}$, $6\,\upmu\mathrm{m}$). The input parameters are taken from the PIC simulations: $T_{h0} = 0.7\,\mathrm{MeV}$, $\theta_h=30^\circ$, $n_{h0}=2n_c$ (resp. $1n_c$) at $d=3\,\upmu\mathrm{m}$ (resp. $6\,\upmu\mathrm{m}$). The saturation exhibited by all curves happen when $x_f(t) \gtrsim D_\perp$. The maximum energies predicted in 3D, $E_{max} \simeq 3.6\,\mathrm{MeV}$ (resp. $3.3\,\mathrm{MeV}$) at $d=3\,\upmu\mathrm{m}$ (resp. $6\,\upmu\mathrm{m}$) closely match the experimental data. Owing to a weaker mitigation factor at large expansion distances, the 2D calculations yield a higher value, $E_{max} \simeq 4.7\,\mathrm{MeV}$, similar at $d=3\,\upmu\mathrm{m}$ and $6\,\upmu\mathrm{m}$. This somewhat fortuitous constancy results from a compensation of the weakened electric field and the longer-lived 1D expansion regime that occurs at $d=6\,\upmu\mathrm{m}$. 



The expansion time defined by $x_f(t_{acc}) = D_\perp$ can be viewed as the effective proton acceleration time by a single pulse, and, consequently, as the maximum time delay for efficient coupling between the two pulses regarding TNSA. For $d=3\,\upmu\mathrm{m}$ (resp. $6\,\upmu\mathrm{m}$), the 3D model gives $t_{acc} \simeq 430\,\mathrm{fs}$ (resp. $\simeq 660\,\mathrm{fs}$). These values qualitatively agree, yet somewhat underestimate the measured $\Delta t_{0,max} \simeq 600\,\rm fs$ (resp. $\simeq 1\,\rm ps$) at $d=3\,\upmu\rm m$ (resp. $6\,\upmu\rm m$). The difference can be due to several factors that are not, or improperly, modelled: the transverse extent of the sheath field may evolve as the recirculating hot electrons diffuse away transversely; 
our model does not take into account the behaviour of the aluminum ions and of the secondary hot-electron source.


\section{Conclusions}
\label{sec:concl}

In this paper, we have analysed TNSA proton beams generated by the interaction of two femtosecond laser pulses with a thin aluminum target. Both experimental and numerical results have shown that, in a $3\,\upmu\rm m$-thick target, the acceleration process can be affected by the second pulse for time delays as long as $\sim 600\,\rm fs$. Plasma expansion is induced by the first pulse, which leads to better absorption and higher efficiency of the second half-pulse. The hot electrons generated by the second pulse then overtake the proton beam accelerated by the first pulse, which can consequently get a boost of its energy from the renewed sheath electric field. This boost will however be limited due to multidimensional dilution of the accelerating fields on the rear side, combined with prior proton propagation. We have also developed a simple `unified' model for TNSA that captures the transition from the isothermal to the adiabatic electron regimes, as well as, in a more qualitative way, the saturation caused by dimensional effects.

The precise experimental control of the time delay between the two pulses and the analysis based on 2D PIC simulations enable us to study the multidimensional effects of plasma expansion on the interaction with the second pulse. In this respect, this study constitutes a step forward compared with the work presented in Ref.~\onlinecite{Markey2010}. We observe similar phenomena with our much shorter laser pulses and thinner targets, indicating that these phenomena are indeed largely reproducible with different parameters. However, our study also points out that the maximum proton energy cannot be increased by simply choosing an appropriate time delay when the laser energy is equally distributed between the two sub-pulses. In that respect, modifying the energy ratio between the two sub-pulses as in Ref.~\onlinecite{Markey2010} should be beneficial.

Most of the numerical results presented here assume the absence of an initial plasma gradient on the front and rear sides of the target, which might not be the case under experimental conditions (it would require an even higher contrast than the one attainable herein). The underlying physics is likely to be changed by the presence of a preplasma before the first pulse, and should be partly responsible for the discrepancy observed between experimental and numerical results. Another contributing factor to this discrepancy are dimensional effects, as the damping of the accelerating field is enhanced for plasma expansion in three dimensions. However, this should not change the main conclusion, as the target will still undergo significant expansion due to the interaction with the first half-pulse. Even with a preplasma initially present on the front side of the target, the second half-pulse should still influence the proton dynamics over a relatively long time span.

\acknowledgments 
This work was supported by the Knut and Alice Wallenberg Foundation and by the Swedish Research Council. The simulations were performed on resources at
Chalmers Centre for Computational Science and Engineering (C3SE) provided by the Swedish National Infrastructure for
Computing (SNIC). EPOCH was developed under UK EPSRC grants EP/G054950/1, EP/G056803/1, EP/G055165/1 and
EP/M022463/1.

\end{document}